\documentclass{article}
\usepackage{spconf,amsmath,graphicx,amssymb,algorithm,algorithmic}

\title{Alternating projections gridless covariance-based estimation for DOA}
\name{Yongsung Park, Peter Gerstoft\thanks{Supported by the Office of Naval Research, Grant No. N00014-18-1-2118.}}
\address{Scripps Institution of Oceanography, University of California San Diego, La Jolla, CA, 92093-0238}
\begin{document}
%
\maketitle
\begin{abstract}
We present a gridless sparse iterative covariance-based estimation method based on alternating projections for direction-of-arrival (DOA) estimation.
The gridless DOA estimation is formulated in the reconstruction of Toeplitz-structured low rank matrix, and is solved efficiently with alternating projections.
The method improves resolution by achieving sparsity, deals with single-snapshot data and coherent arrivals, and, with co-prime arrays, estimates more DOAs than the number of sensors.
We evaluate the proposed method using simulation results focusing on co-prime arrays.
\end{abstract}
\section{Introduction}
\label{sec:intro}
\vspace*{-1mm}
Direction-of-arrival (DOA) estimation is localizing several sources arriving at an array of sensors.
It is an important problem in a wide range of applications, including radar, sonar, etc.
Compressive sensing (CS) based DOA estimation, which promotes sparse solutions, has advantages over traditional DOA estimation methods.~\cite{malioutov05asparse,gerstoft15multicsbeam}
DOAs exist on a continuous angular domain, and gridless CS can be employed.~\cite{xenaki15gridfree,park2018grid,yang18DOAreview}
We propose a gridless sparsity-promoting DOA estimation method and apply it to co-prime arrays, which can resolve more sources than the number of sensors.

CS-based DOA estimation exploits the framework of CS, which promotes sparse solutions, for DOA estimation and has a high-resolution capability, deals with single-snapshot data, and performs well with coherent arrivals.~\cite{gerstoft15multicsbeam,xenaki15gridfree}
To estimate DOAs in a continuous angular domain, non-linear estimation of the DOAs is linearized by using a discretized angular-search grid of potential DOAs (``steering vectors'').
Grid-based CS has a basis mismatch problem~\cite{tang13csoffthegrid,chi11basismismatch} when true DOAs do not fall on the angular-search grid.
To overcome the basis mismatch, gridless CS~\cite{chi20anmreview,wang2019superresolution} has been utilized for DOA estimation.~\cite{xenaki15gridfree,park2018grid,yang18DOAreview,raj19icassp,semper18icassp}

Gridless SPICE (GLS)~\cite{yang18DOAreview,yang14glspice}, one of the off-grid sparse methods, is a gridless version of sparse iterative covariance-based estimation (SPICE)~\cite{stoica11spice1}.
GLS re-parameterizes the data covariance, or sample covariance matrix (SCM), using a positive semi-definite (PSD) Toeplitz matrix, and finds the lowest rank Toeplitz matrix which fits the SCM.
The Toeplitz-structured SCM is related to a Fourier-series, which is composed of harmonics.~\cite{zhu2017asymptotic,romero2013wideband}
GLS-based DOA estimation retrieves DOA-dependent harmonics from the SCM parameter.~\cite{yang18DOAreview}
The GLS solver uses a semi-definite programming problem (SDP), which is infeasible in practice for high-dimensional problems.

Alternating projections (AP) algorithm~\cite{boyd2004convex,dattorro2010convex} has been introduced to solve matrix completion~\cite{cai19ap_matcomp,jiang17ap_matcomp,lewis08ap_lrp,netrapalli14ap_matcomp} and structured low rank matrix recovery~\cite{cho16ap_lowrank,condat15ap_lowrank} and consists of projecting a matrix onto the intersection of a linear subspace and a nonconvex manifold.
Atomic norm minimization (ANM)~\cite{tang13csoffthegrid,chi20anmreview} solves gridless CS and is equivalent to a recovery of a Toeplitz-structured low rank matrix~\cite{wu17toeplitzdoa}.
AP based on ANM has been applied to gridless CS for DOA estimation.~\cite{wagner2019gridless,wagner2020gridless}

We propose AP-based GLS for gridless CS for DOA estimation.
GLS reconstructs a DOA-dependent SCM matrix, which is a Toeplitz-structured low rank matrix and has a PSD matrix in its constraint.
AP-GLS solves the reconstruction of the Toeplitz-structured low rank matrix by using a sequence of projections onto the following sets: Toeplitz set, rank-constraint set, and PSD set.

Co-prime arrays are introduced for DOA estimation and offer the capability of identifying more sources than the number of sensors.~\cite{pal10coprime}
Sparse Bayesian learning (SBL) deals with co-prime arrays without constructing a co-array based covariance matrix and shows accurate DOAs identifying more sources than the number of sensors.~\cite{santosh18,santosh18sam}
We apply AP-GLS to co-prime arrays and show that AP-GLS with co-prime arrays estimates more DOAs than the number of sensors.

We study the performance of AP-GLS with co-prime arrays for single- and multiple-snapshot data, incoherent and coherent sources, and when the number of sources exceeds the number of sensors.

\vspace*{-1mm}
\section{Signal model and co-prime array}
\label{sec:format}
\vspace*{-1mm}
\subsection{Signal model}
\vspace*{-2mm}
We consider $K$ narrowband sources for $L$ snapshot data with complex signal amplitude $s_{k,l} \in \mathbb{C}$, $k=1,\dotsc,K$, $l=1,\dotsc,L$.
The sources have stationary DOAs for $L$ snapshots $\theta_{k} \in \Theta \triangleq [-90^\circ, 90^\circ)$, $k=1,\dotsc, K$ in the far-field of a linear array with $M$ sensors.
The observed data $\mathbf{Y} \in \mathbb{C}^{M \times L}$ is modeled as
\begin{equation}
\mathbf{Y} = \sum_{k=1}^{K} \mathbf{a}(\theta_k) \mathbf{s}_{k:} +\mathbf{E} 
= \sum_{k=1}^{K} c_k \mathbf{a}(\theta_k) \boldsymbol{\phi}_{k:} +\mathbf{E},
\end{equation}
where $\mathbf{s}_{k:} = [ s_{k,1} \dotsc s_{k,L} ] \in \mathbb{C}^{1 \times L}$, $c_k = \lVert \mathbf{s}_{k:} \rVert _2 >0$, $\boldsymbol{\phi}_{k:}=c_k^{-1} \mathbf{s}_{k:} \in \mathbb{C}^{1 \times L}$ with $\lVert \boldsymbol{\phi}_{k:} \rVert_2 = 1$, $\mathbf{E} \in \mathbb{C}^{M \times L}$ is the measurement noise, and $\mathbf{a}(\theta_k) \in \mathbb{C}^M$ is the steering vector.
The steering vector is given by ($\lambda$ is the signal wavelength and $d_m$ is the distance from sensor~$1$ to sensor~$m$)
\begin{equation}
\mathbf{a}(\theta_k) = \Big[ 1 \ e^{-j \frac{2 \pi}{\lambda} d_2 \sin \theta_k} \ \dotsc \ e^{-j \frac{2 \pi}{\lambda} d_M \sin \theta_k} \Big]^\mathsf{T}.
\end{equation}

\vspace*{-4mm}
\subsection{Co-prime array}
\vspace*{-2mm}
Consider the sensor positions in an array is given by $d_m = \delta_m d$, $m=1,\dotsc,M$, where the integer $\delta_m$ is the normalized sensor location of $m$th sensor and $d$ is the minimum sensor spacing.
A uniform linear array (ULA) is composed of uniformly spaced sensors with $\boldsymbol{\delta} \hspace*{-.3mm}=\hspace*{-.3mm} [0 \ 1 \dotsc M-1]^\mathsf{T}$ and $d =\hspace*{-.3mm} \lambda / 2$.

A co-prime array involves two ULAs with spacing $M_1 d$ and $M_2 d$ where $M_1$ and $M_2$ are co-prime, i.e., their greatest common divisor is 1.~\cite{pal10coprime}
A co-prime array consists of a ULA with $\boldsymbol{\delta} = [0 \ M_2 \dotsc (M_1-1)M_2]^\mathsf{T}$ and a ULA with $\boldsymbol{\delta} = [M_1 \ 2 M_1 \dotsc (2 M_2-1) M_1]^\mathsf{T}$, a total of $M_1 + 2 M_2 - 1$ sensors.

We used a 16-sensor ULA with $\boldsymbol{\delta} = [0 \ 1 \dotsc 15]^\mathsf{T}$ and a 8-sensor co-prime array with $M_1 = 5$ and $M_2 = 2$, i.e., $\boldsymbol{\delta} = [0 \ 2 \ 4 \ 5 \ 6 \ 8 \ 10 \ 15]^\mathsf{T}$.

\section{Alternating\hspace*{-.3mm} projections\hspace*{-.3mm} gridless\hspace*{-.3mm} SPICE}
\label{sec:apgls}
\vspace*{-1mm}

Consider the ULA case and assume incoherent sources. (GLS is robust to source correlations.~\cite{yang18DOAreview,yang14glspice,stoica11spice1})
In the noiseless case, the SCM $\mathbf{R}^\ast \in \mathbb{C}^{M \times M}$ is given by
\begin{equation}
\mathbf{R}^\ast = \frac{1}{L} \mathbf{Y}^\ast \mathbf{Y}^{\ast \mathsf{H}} = \sum_{k=1}^{K} p_k \mathbf{a}(\theta_k) \mathbf{a}^\mathsf{H}(\theta_k)
\end{equation}
where $\mathbf{Y}^\ast$ is noise-free data and $p_k > 0$, $k=1,\dotsc,K$ is the power of sources, i.e., $p_k = c_k^2$.
The SCM $\mathbf{R}^\ast$ is a (Hermitian) Toeplitz matrix,
\begin{equation}
{\mathbf{R}^\ast} = \mathrm{Toep} (\mathbf{r}) =
\begin{bmatrix}
{r}_{1} 			& {r}_{2}				& \cdots 	& {r}_{M} 		\\
{r}_{2}^{\mathsf{H}} 	& {r}_{1} 				& \cdots 	& {r}_{M-1} 	\\
\vdots 			& \vdots 				& \ddots 	& \vdots 		\\
{r}_{M}^{\mathsf{H}} 	& {r}_{M-1}^{\mathsf{H}} 	& \cdots 	& {r}_{1} 		\\
\end{bmatrix},
\end{equation}
where $\mathbf{r} \in \mathbb{C}^M$.
Moreover, ${\mathbf{R}^\ast}$ is PSD and has rank $K$.
A PSD Toeplitz matrix of rank $K < M$ can be uniquely decomposed (Vandermonde decomposition)~\cite{yang18DOAreview,tang13csoffthegrid,steffens18vanderm} as
\begin{equation}
\mathbf{R}^\ast = \sum_{k=1}^{K} p_k \hspace*{.3mm}\mathbf{a}(\theta_k) \hspace*{.3mm} \mathbf{a}^\mathsf{H}(\theta_k)
= \mathbf{A} \hspace*{.3mm}\mathrm{diag}(\mathbf{p}) \mathbf{A}^\mathsf{H},
\label{eq:vandermonde}
\end{equation}
where $\mathbf{A} = [\mathbf{a}(\theta_1) \ \dotsc \ \mathbf{a}(\theta_K)] \in \mathbb{C}^{M \times K}$.

GLS uses a SCM-related parameter $\mathbf{R} \in \mathbb{C}^{M \times M}$, which is a rank-$K$ PSD Toeplitz matrix, and fits the parameter $\mathbf{R}$ to SCM $\tilde{\mathbf{R}} = \mathbf{Y} \mathbf{Y}^\mathsf{H} / L \in \mathbb{C}^{M \times M}$.
The covariance fitting is implemented, in the case of $L \geq M$ whenever $\tilde{\mathbf{R}}$ is non-singular, by minimizing the criterion,~\cite{yang18DOAreview,yang14glspice,stoica11spice1}
\begin{equation}
\left\lVert  \mathbf{R}^{-\frac{1}{2}} \left( \tilde{\mathbf{R}}-\mathbf{R} \right) \tilde{\mathbf{R}}^{-\frac{1}{2}} \right\rVert_\mathrm{F}^2.
\end{equation}
In the case of $L < M$, when $\tilde{\mathbf{R}}$ is singular, the following criterion is used instead,~\cite{yang18DOAreview,yang14glspice}
\begin{equation}
\left\lVert  \mathbf{R}^{-\frac{1}{2}} \left( \tilde{\mathbf{R}}-\mathbf{R} \right) \right\rVert_\mathrm{F}^2
= \mathrm{tr} \big( \tilde{\mathbf{R}} \mathbf{R}^{-1} \tilde{\mathbf{R}} \big) + \mathrm{tr} \left( \mathbf{R} \right) - 2 \mathrm{tr} ( \tilde{\mathbf{R}} ).
\end{equation}
GLS is achieved using the following optimization,
%
\begin{align}
& \underset{\mathbf{R}}{\text{min}} \ \
\mathrm{tr} \big( \tilde{\mathbf{R}} \mathbf{R}^{-1} \tilde{\mathbf{R}} \big) + \mathrm{tr} ( \mathbf{R} )
& & \hspace*{-1mm} \text{subject to} \ \ 
\mathbf{R} \succeq 0 \nonumber \\ 
& \Leftrightarrow \underset{\mathbf{R},\mathbf{Z}}{\text{min}} \ \
\mathrm{tr} ( \mathbf{Z} ) + \mathrm{tr} ( \mathbf{R} ) 
& & \hspace*{-1mm} \text{subject to} \
\begin{cases}
    \mathbf{R} \succeq 0\\
    \mathbf{Z} \succeq \tilde{\mathbf{R}} \mathbf{R}^{-1} \tilde{\mathbf{R}},
\end{cases}
%
\nonumber \\
& \Leftrightarrow \underset{\mathbf{R},\mathbf{Z}}{\text{min}} \ \
\mathrm{tr} ( \mathbf{Z} ) + \mathrm{tr} ( \mathbf{R} ) 
& & \hspace*{-1mm} \text{subject to} \
\begin{bmatrix}
\mathbf{R} 		& \tilde{\mathbf{R}} \\
\tilde{\mathbf{R}} 	& \mathbf{Z}
\end{bmatrix}
\succeq 0,
\label{eq:GLSopt}
\end{align}
where $\mathbf{R} \succeq 0$ denotes $\mathbf{R}$ is a PSD matrix and $\mathbf{Z} \in \mathbb{C}^{M \times M}$ is a free variable.
Consider the case of $\mathbf{R} = \mathbf{R}^\ast$, then 
%
$\mathrm{tr} ( \mathbf{R} ) = M \sum_{k=1}^{K} p_k$.
%
Defining $\mathrm{tr} (\mathbf{Z}) = M \sum_{k=1}^{K} p_k$, the objective in~\eqref{eq:GLSopt}, divided by $2M$, equals,
\begin{equation}
\frac{1}{2M} \mathrm{tr} ( \mathbf{R} ) + \frac{1}{2M} \mathrm{tr} ( \mathbf{Z} ) = \sum_{k=1}^{K} p_k.
\end{equation}
Note that, in ANM,~\cite{tang13csoffthegrid,chi20anmreview} minimizing $\sum_{k=1}^{K} p_k = \sum_{k=1}^{K} c_k^2$ is equivalent to minimizing the atomic norm,
\begin{equation}
\left \lVert \mathbf{Y}^\ast \right \rVert_\mathcal{A} \hspace*{-.3mm} = \hspace*{-.3mm}
\underset{c_k,\theta_k,\boldsymbol{\phi}_{k:}}{\text{inf}}
\hspace*{-.7mm}
\left\{ \sum_{k=1}^{K} c_k: \mathbf{Y}^\ast \hspace*{-.3mm} = \sum_{k=1}^{K} c_k \hspace*{.3mm} \mathbf{a}(\theta_k) \boldsymbol{\phi}_{k:} \right\} \hspace*{-.7mm}.
\end{equation}
The atomic norm is a convex relaxation of the atomic $l_0$ norm,~\cite{tang13csoffthegrid}
\begin{equation}
\left \lVert \mathbf{Y}^\ast \right \rVert_{\mathcal{A},0} \hspace*{-.3mm} = \hspace*{-.3mm}
\underset{c_k,\theta_k,\boldsymbol{\phi}_{k:}}{\text{inf}} 
\hspace*{-.7mm}
\left\{ K : \mathbf{Y}^\ast \hspace*{-.3mm} = \sum_{k=1}^{K} c_k \hspace*{.3mm} \mathbf{a}(\theta_k) \boldsymbol{\phi}_{k:} \right\}
\hspace*{-.7mm}.
\end{equation}
Minimizing the atomic $l_0$ norm is equivalent to minimizing rank of $\mathbf{R}^\ast = \mathbf{Y}^\ast \mathbf{Y}^{\ast \mathsf{H}}/L$.~\cite{yang18DOAreview,tang13csoffthegrid}
Summarizing, the term $\mathrm{tr} ( \mathbf{R} ) = \sum_{k=1}^{K} p_k$ is the nuclear norm, used as a convex relaxation of $\mathrm{rank}( \mathbf{R} )$.

By using the rank minimization in \eqref{eq:GLSopt}, the resulting optimization is as follows,
\begin{equation}
\underset{\mathbf{R},\mathbf{Z}}{\text{min}}
\ \ \mathrm{rank} ( \mathbf{R} ) \quad
\text{subject to} \ \
\begin{bmatrix}
\mathbf{R} 		& \tilde{\mathbf{R}} \\
\tilde{\mathbf{R}} 	& \mathbf{Z}
\end{bmatrix}
\succeq 0.
\label{eq:GLSrankopt}
\end{equation}

For the coprime array, we use the row-selection matrix $\mathbf{\Gamma}_{\Omega} \in \{ 0,1 \}^{M \times M_\Omega}$, i.e.,\vspace{-.6mm}
\begin{equation}\vspace{-.6mm}
\mathbf{Y}_\Omega = \mathbf{\Gamma}_\Omega \mathbf{Y} \ \text{or} \ \mathbf{Y} = \mathbf{\Gamma}_\Omega^\dagger \mathbf{Y}_\Omega,
\end{equation}
where $\mathbf{Y}$ is data of full $M$-element ULA and the Moore-Penrose pseudo-inverse $\mathbf{\Gamma}_\Omega^\dagger$.
The optimization for the coprime array is given as,\vspace{-.6mm}
\begin{equation}\vspace{-.6mm}
\underset{\mathbf{R},\mathbf{Z}}{\text{min}}
\ \ \mathrm{rank} ( \mathbf{R} ) \quad
\text{subject to} \ \
\begin{bmatrix}
\mathbf{R}_\Omega 		& \tilde{\mathbf{R}}_\Omega \\
\tilde{\mathbf{R}}_\Omega 	& \mathbf{Z}
\end{bmatrix}
\succeq 0,
\label{eq:GLSrankoptS}
\end{equation}
where $\tilde{\mathbf{R}}_\Omega = \mathbf{Y}_\Omega \mathbf{Y}_\Omega^\mathsf{H} / L \in \mathbb{C}^{M_\Omega \times M_\Omega}$ and $\mathbf{R}_\Omega = \mathbf{\Gamma}_\Omega \mathbf{R} \mathbf{\Gamma}_\Omega^\mathsf{T} \in \mathbb{C}^{M_\Omega \times M_\Omega}$.
To minimize $\mathrm{rank} (\mathbf{R})$, $\mathbf{R}$ is calculated,\vspace{-1mm}
\begin{equation}\vspace{-1mm}
\mathbf{R} = \mathbf{\Gamma}_\Omega^\dagger \mathbf{R}_\Omega {( \mathbf{\Gamma}_\Omega^\dagger )}^\mathsf{T}.
\label{eq:rBack}
\end{equation}

\section{Alternating projections}
\label{sec:typestyle}
\vspace*{-1mm}
We suggest alternating projections to reconstruct Toeplitz-structured low rank matrix in \eqref{eq:GLSrankopt} and \eqref{eq:GLSrankoptS}.
AP-GLS involves the following sets: Toeplitz set, positive semi-definite (PSD) set, and rank-constraint set.

\vspace*{-3mm}
\subsection{Projection onto the Toeplitz set} \vspace*{-2mm}
The SCM-related parameter $\mathbf{R}$ is a Toeplitz matrix, and the projection of $\mathbf{R}$ onto the Toeplitz set $\mathcal{T}$ is implemented by finding the closest Toeplitz matrix,~\cite{cho16ap_lowrank,eberle03ap_toeplitz}
\begin{eqnarray}
P_{\mathcal{T}}(\mathbf{R}) = \mathrm{Toep} (\mathbf{r}), \quad\quad\quad\quad \ \
\label{eq:pToep}
\\
r_m = \frac{1}{2(M-m)} \sum_{i=1}^{M-m} {R}_{i,i+m-1} + {R}_{i+m-1,i}^{\mathsf{H}}.
\label{b}
\end{eqnarray}
Note that, $m$th component of $\mathbf{r} \in \mathbb{C}^{M}$ is obtained by averaging $m$th diagonal and the conjugate diagonal components.

\vspace*{-3mm}
\subsection{Projection onto the PSD set} \vspace*{-2mm}
The constraints \eqref{eq:GLSrankopt} and \eqref{eq:GLSrankoptS} include PSD matrices, which is obtained by projecting the matrix in the constraint onto the PSD set $\mathcal{P}$, defined by the PSD cone.
The projection of a (Hermitian) matrix $\mathbf{S}$ onto the PSD set is achieved from the eigen-decomposition $\mathbf{S}=\sum_{i=1}^{2M} \mu_i \mathbf{q}_i \mathbf{q}_i^\mathsf{H}$,~\cite{boyd2004convex,dattorro2010convex}
\begin{equation}
P_{\mathcal{P}}(\mathbf{S}) = \sum_{i=1}^{2M} \mathrm{max} \{ 0,\mu_i \} \mathbf{q}_i \mathbf{q}_i^\mathsf{H}.
\label{eq:pPsd}
\end{equation}
%

\vspace*{-3mm}
\subsection{Projection onto the rank-constraint set} \vspace*{-2mm}
The objectives \eqref{eq:GLSrankopt} and \eqref{eq:GLSrankoptS} include rank-constraints.
Consider the case of rank-$K$ matrix $\mathbf{R}$. 
The projection of $\mathbf{R}$ onto the rank-constraint set $\mathcal{R}$ is achieved from the singular value decomposition and taking the $K$-largest singular values,~\cite{jiang17ap_matcomp,condat15ap_lowrank}
\begin{equation}
P_{\mathcal{R}}(\mathbf{R}) = \sum_{k=1}^{K} \sigma_k \mathbf{u}_k \mathbf{v}_k^\mathsf{H},
\label{eq:pRank}
\end{equation}
where $\sigma_k$, $\mathbf{u}_k \hspace*{-1mm} \in \hspace*{-.6mm} \mathbb{C}^{M}$, $\mathbf{v}_k \hspace*{-1mm} \in \hspace*{-.6mm} \mathbb{C}^{M}$, $k \hspace*{-.6mm}=\hspace*{-.6mm} 1,\dotsc,K$, are the $K$-largest singular values and the corresponding left and right singular vectors.
We remark that $\mathbf{R}$ is an SCM, thus the eigen-decomposition and the singular value decomposition result in the same results.

\vspace*{-3mm}
\subsection{Alternating projections} \vspace*{-2mm}
Initialized parameters $\mathbf{R}$ and $\mathbf{Z}$ form $\mathbf{S}$, which is PSD, i.e., $\mathbf{S} = P_{\mathcal{P}}(\mathbf{S})$~\eqref{eq:pPsd}.
$\mathbf{R}$ is obtained from $\mathbf{S}$, $\mathbf{R} = \mathbf{\Gamma}_\Omega^\dagger \mathbf{S}(1:M,1:M) {( \mathbf{\Gamma}_\Omega^\dagger )}^\mathsf{T}$~\eqref{eq:rBack}, and the projection $P_{\mathcal{R}}(\mathbf{R})$~\eqref{eq:pRank} is carried out to make $\mathbf{R}$ be $K$-rank.
The projection $P_{\mathcal{T}}(\mathbf{R})$~\eqref{eq:pToep} is followed for a Toeplitz structure.
Submatrix $\mathbf{\Gamma}_\Omega \mathbf{R} \mathbf{\Gamma}_\Omega^\mathsf{T}$ is implemented in $\mathbf{S}$.
AP-GLS iterates the projections until it converges to a solution.
AP-GLS is summarized in Algorithm 1.

\vspace*{-3mm}
\subsection{DOA retrieval} \vspace*{-2mm}
DOAs $\theta_k$, $k = 1,\dotsc,K$, are recovered by the Vandermonde decomposition \eqref{eq:vandermonde} for the rank-$K$ PSD Toeplitz matrix $\mathbf{R}$.~\cite{yang18DOAreview,tang13csoffthegrid,chi20anmreview}
The Vandermonde decomposition is computed efficiently via root-MUSIC~\cite{wagner2020gridless}:  \par
\setlength{\leftskip}{.5cm}
\noindent\hangindent=0.3cm 1. Perform the eigen-decomposition in signal- and noise-subspace, i.e., $\mathbf{R} = \mathbf{U}_S \boldsymbol{\Lambda}_S \mathbf{U}_S^\mathsf{H} + \mathbf{U}_N \boldsymbol{\Lambda}_N \mathbf{U}_N^\mathsf{H}$.

\noindent\hangindent=0.3cm 2. Compute the root-MUSIC polynomial $\mathcal{Q}(z) \hspace*{-.6mm}=\hspace*{-.5mm} \mathbf{a}^\mathsf{T}(1/z)$ $\mathbf{U}_N \mathbf{U}_N^\mathsf{H} \mathbf{a}(z)$, where 
$\mathbf{a} \hspace*{-.6mm}=\hspace*{-.6mm} [1,z,\dotsc,z^{M-1}]^\mathsf{T}$ \hspace*{-.6mm}and
$z \hspace*{-.6mm}=\hspace*{-.6mm} e^{ - j (2 \pi / \lambda) d \sin \theta}$\hspace*{-.3mm}.

\noindent\hangindent=0.3cm 3. Find the roots of $\mathcal{Q}(z)$ and choose the $K$ roots that are inside the unit circle and closest to the unit circle, i.e., $\hat{z}_i$, $i=1,\dotsc,K$.

\noindent\hangindent=0.3cm 4. DOA estimates are recovered, i.e., $\hat{\theta}_i = -\sin^{-1}( \frac{\lambda \angle \hat{z}_i}{2 \pi d}  )$, $i=1,\dotsc,K$.

\setlength{\leftskip}{0pt}

 \begin{algorithm}[t]
 \caption{AP-GLS}
 \begin{algorithmic}[1]
  \STATE Input: $\mathbf{Y} \in \mathbb{C}^{M \times L}$, $K$, $\mathbf{\Gamma}_\Omega$
  \STATE Parameters: $\epsilon_\mathrm{min} = 10^{-3}$
  \STATE Initialization: $\mathbf{R} \in \mathbb{C}^{M \times M}$, $\mathbf{Z} \in \mathbb{C}^{M \times M}$ with uniformly $(0,1)$ distributed random for real and imaginary part.
  \STATE $\mathbf{R}^\text{old} = \mathbf{R}$, $\mathbf{Z}^\text{old} = \mathbf{Z}$
  \WHILE {$\lVert \mathbf{S}-\mathbf{S}^\text{old} \rVert_\mathrm{F} < \epsilon_\mathrm{min}$}
  \STATE $\mathbf{S} = 
  		\begin{bmatrix}
		\mathbf{\Gamma}_\Omega \mathbf{R}^\text{old} \mathbf{\Gamma}_\Omega^\mathsf{T}&\tilde{\mathbf{R}}_\Omega \\
		\tilde{\mathbf{R}}_\Omega 	& \mathbf{Z}^\text{old}
		\end{bmatrix}$
  \STATE PSD projection: $\mathbf{S} = P_{\mathcal{P}}(\mathbf{S})$~\eqref{eq:pPsd}
  \STATE $\mathbf{R} = \mathbf{\Gamma}_\Omega^\dagger \mathbf{S}(1:M,1:M) {( \mathbf{\Gamma}_\Omega^\dagger )}^\mathsf{T}$~\eqref{eq:rBack}
  \STATE Rank-constraint projection: $\mathbf{R} = P_{\mathcal{R}}(\mathbf{R})$~\eqref{eq:pRank}
  \STATE Toeplitz projection: $\mathbf{R} = P_{\mathcal{T}}(\mathbf{R})$~\eqref{eq:pToep}
  \STATE $\mathbf{S}(1:M,1:M) = \mathbf{\Gamma}_\Omega \mathbf{R} \mathbf{\Gamma}_\Omega^\mathsf{T}$
  \STATE $\mathbf{R}^\text{old} = \mathbf{R}$, $\mathbf{Z}^\text{old} = \mathbf{S}(M+1:2M,M+1:2M)$
  \ENDWHILE
 \STATE Output: $\mathbf{R}$
 \end{algorithmic}
 \end{algorithm}


\begin{figure}[t]

\begin{minipage}[b]{1.0\linewidth}
  \centering
  \centerline{\includegraphics[width=8.0cm]{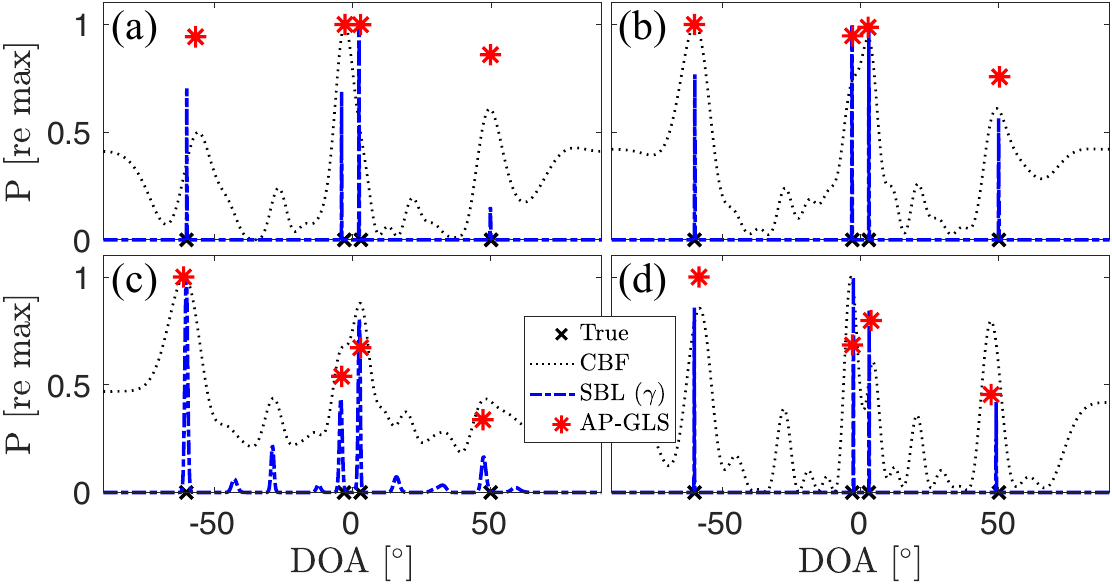}}
\vspace{-3mm}
\end{minipage}
\caption{DOA estimation from $L$ snapshots for $K\hspace*{-.9mm}=\hspace*{-.3mm}4$ sources with an $M\hspace*{-.9mm}=\hspace*{-.3mm}8$ co-prime array. CBF, SBL, and AP-GLS for (a) incoherent sources with SNR 20~dB and one snapshot, $L\hspace*{-.7mm}=\hspace*{-.3mm}1$, (b) SNR 20~dB and $L\hspace*{-.7mm}=\hspace*{-.3mm}20$, (c) SNR 0~dB and $L\hspace*{-.7mm}=$ $\hspace*{-.3mm}20$, and (d) for coherent sources with SNR 20~dB and $L\hspace*{-.7mm}=\hspace*{-.3mm}20$.}
\label{fig:fig1}
\end{figure}

\begin{figure}[t]

\begin{minipage}[b]{1.0\linewidth}
  \centering
  \centerline{\includegraphics[width=8.0cm]{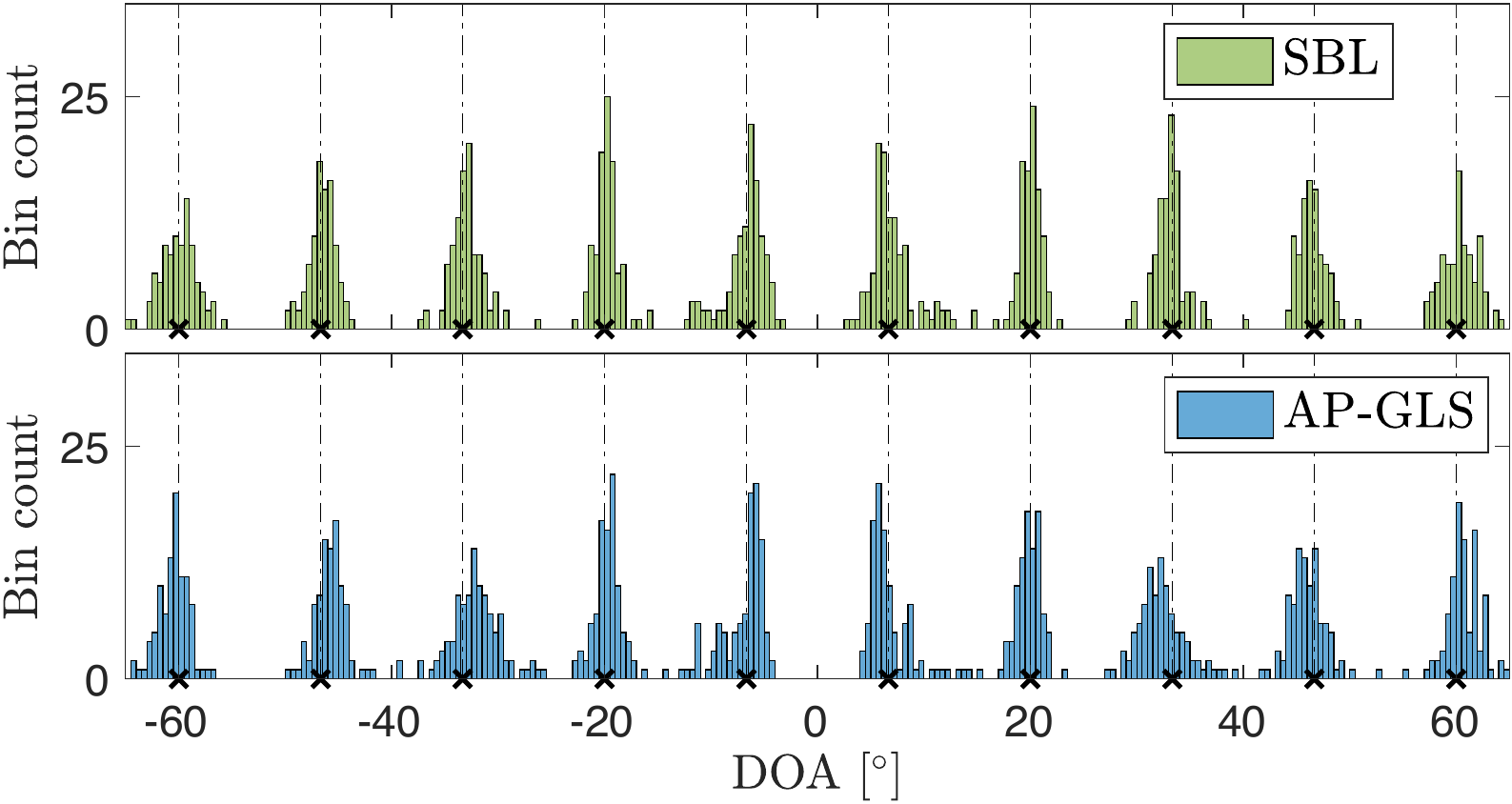}}
\vspace{-3mm}
\end{minipage}
\caption{Histogram of the estimated DOAs of SBL and AP-GLS for $K=10$ sources with an $M=8$ co-prime array for 100 trials. ($M<K$)}
\label{fig:fig2}
\end{figure}

\begin{figure}[t]

\begin{minipage}[b]{1.0\linewidth}
  \centering
  \centerline{\includegraphics[width=8.0cm]{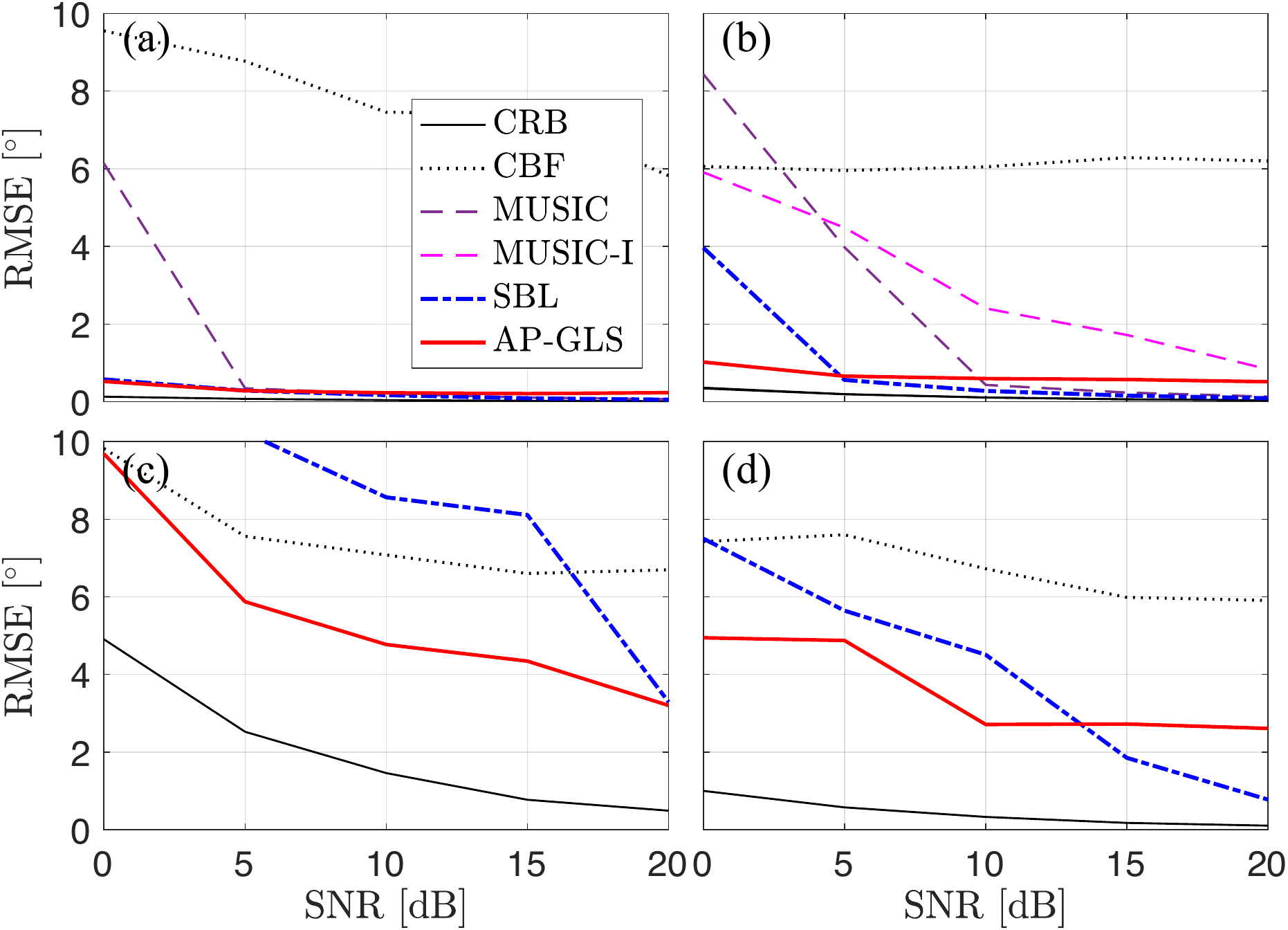}}
\vspace{-3mm}
\end{minipage}
\caption{RMSE~[$^\circ$] comparison versus SNR. Each RMSE is averaged over 100 trials. $K\hspace*{-.9mm}=\hspace*{-.3mm}4$ incoherent sources for (a) an $M\hspace*{-.9mm}=\hspace*{-.3mm}16$ ULA with $L\hspace*{-.7mm}=\hspace*{-.3mm}20$, (b) an $M\hspace*{-.9mm}=\hspace*{-.3mm}8$ co-prime array with $L\hspace*{-.7mm}=\hspace*{-.3mm}20$, (c) $L\hspace*{-.7mm}=\hspace*{-.3mm}1$, and (d) coherent sources for a co-prime array with $L\hspace*{-.7mm}=\hspace*{-.3mm}20$.}
\label{fig:fig3}
\end{figure}

\begin{figure}[t]

\begin{minipage}[b]{1.0\linewidth}
  \centering
  \centerline{\includegraphics[width=8.0cm]{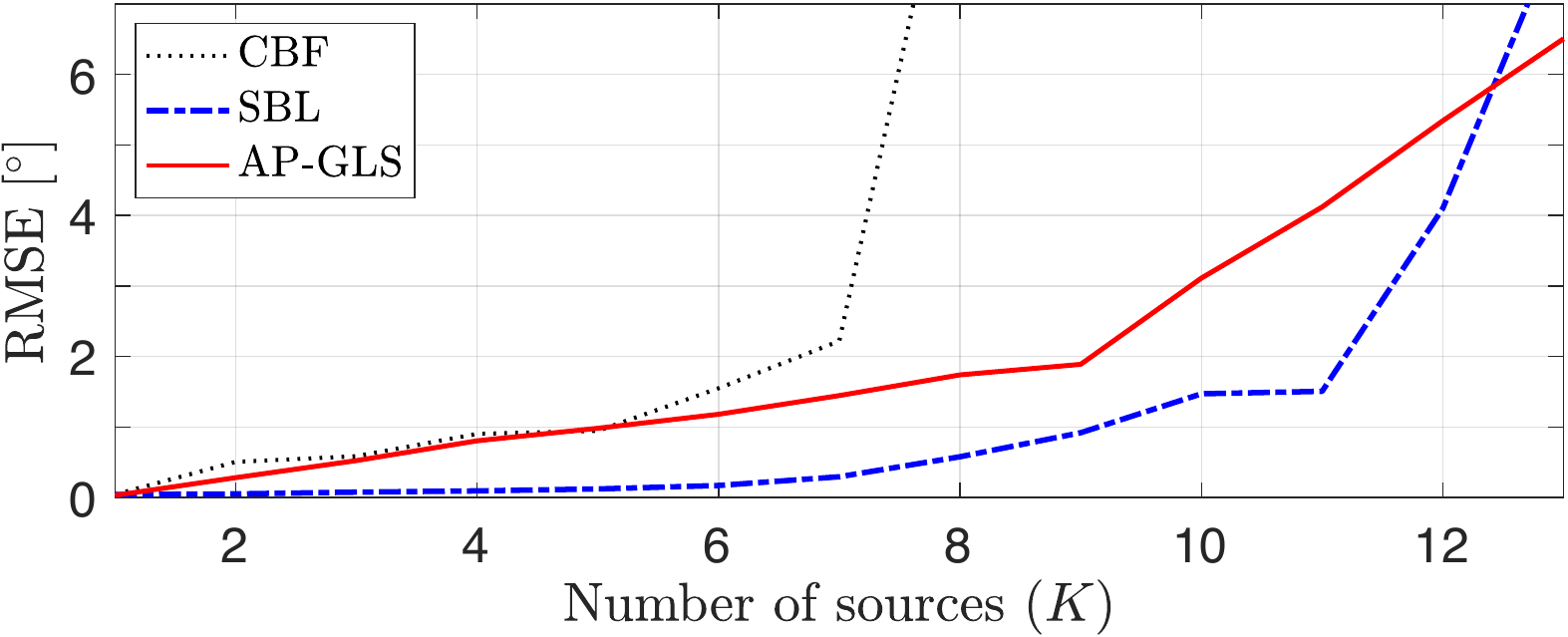}}
\vspace{-3mm}
\end{minipage}
\caption{RMSE~[$^\circ$] comparison versus $K$. Each RMSE is averaged over 100 trials. An $M\hspace*{-.9mm}=\hspace*{-.3mm}8$ co-prime array is used with $L\hspace*{-.7mm}=$ $\hspace*{-.3mm}20$ and SNR 20~dB.}
\label{fig:fig4}
\end{figure}

\section{Simulation results}
\label{sec:simul}
\vspace*{-1mm}
We consider a ULA with $M\hspace*{-.7mm}=\hspace*{-.7mm}16$ elements, half-wavelength spacing, and a co-prime array with $M\hspace*{-.7mm}=\hspace*{-.7mm}8$ elements with $M_1 = 5$ and $M_2 = 2$ (elements 1, 3, 5, 6, 7, 9, 11, 16).

The signal-to-noise ratio (SNR) is defined,
%
$\text{SNR}=10$ $\log_{10}[\mathbb{E} \{ \lVert \mathbf{A} \mathbf{s}_{l} \rVert \}_2^2 / \mathbb{E} \{ \lVert \mathbf{e}_{l} \rVert \}_2^2]$,
%
where $\mathbf{s}_{l} \hspace*{-.7mm}\in\hspace*{-.7mm} \mathbb{C}^K$ and $\mathbf{e}_{l} \hspace*{-.7mm}\in\hspace*{-.7mm} \mathbb{C}^M$\hspace*{-.7mm}, $l\hspace*{-.7mm}=\hspace*{-.7mm}1,\dotsc,L$, are the source amplitude and the measurement noise for the $l$th snapshot.
The root mean squared error (RMSE) is,
%
$\text{RMSE} =\hspace*{-.7mm} \sqrt{  \mathbb{E} \left[ \frac{1}{K} \sum_{k=1}^{K} \left( \hat{\theta}_{k} - {\theta}_{k} \right)^2 \right]  }$,
%
where $\hat{\theta}_{k}$ and ${\theta}_{k}$ represent estimated and true DOA of the $k$th source.

We consider an $M\hspace*{-.7mm}=\hspace*{-.7mm}8$ co-prime array and $K\hspace*{-.7mm}=\hspace*{-.7mm}4$ stationary sources at DOAs $[-60,-3,3,50]^\circ$ with snapshot-varying magnitudes $[12,20]$~dB in four scenarios, see Fig.~1.
Conventional beamforming (CBF), SBL~\cite{santosh18,santosh18sam}, and AP-GLS are compared.
CBF cannot distinguish close two DOAs $[-3,3]^\circ$.
AP-GLS solves the single snapshot case and resolve the close arrivals.
We also consider the DOA performance with a coherent sources due to multipath arrivals.
AP-GLS still shows accurate DOAs.

Co-prime arrays can estimate more sources than the number of sensors, see Fig.~2.
We consider the same co-prime array and $K=10$ stationary sources uniformly distributed in $[-60,60]^\circ$, with $L=20$ and SNR 20~dB.
The histogram shows the distribution of the DOA estimates of AP-GLS.

The DOA performance is evaluated with the RMSE versus SNR, see Fig.~3.
RMSE larger than 10 times the median is outlier and eliminated.
We consider $K=4$ sources, same as in Fig.~1 but with equal strengths.
Cram\'{e}r-Rao bound (CRB)~\cite{stoica89crb}, MUSIC and MUSIC with co-array interpolation (MUSIC-I)~\cite{liu15comusic} are also compared.
Compared to full ULA cases, AP-GLS has a bounded error even with high SNR, which is come from the fact that $\mathbf{R}$ is recovered from its submatrix $\mathbf{\Gamma}_\Omega \mathbf{R} \mathbf{\Gamma}_\Omega^\mathsf{T}$.

The DOA performance is evaluated with the RMSE versus number of sources $K$, see Fig.~4.
We consider the same co-prime array and for each case, $K$ equal strength sources are generated randomly in $[-65,65]^\circ$, with $L=20$ and SNR 20~dB.
AP-GLS has higher estimation accuracy than CBF and estimating more sources than the number of sensors.

\section{Conclusion}
\label{sec:conclusion}
\vspace*{-1mm}
We introduced alternating projections based gridless sparse iterative covariance-based estimation for direction-of-arrival estimation that is gridless and promotes sparse solutions.
Numerical evaluations indicated that the proposed method shows a favorable performance even with single-snapshot data and coherent arrivals.
For co-prime array data, the proposed algorithm resolved more sources than the number of sensors.

\vfill\pagebreak

\small
\bibliographystyle{IEEEbib}
\bibliography{refs_ICASSP2021}

\end{document}